\documentclass{article}

\usepackage{arxiv}

\usepackage[utf8]{inputenc} 
\usepackage[T1]{fontenc}    
\usepackage{hyperref}       
\usepackage{url}            
\usepackage{booktabs}       
\usepackage{amsfonts}       
\usepackage{nicefrac}       
\usepackage{microtype}      
\usepackage{cleveref}       
\usepackage{lipsum}         
\usepackage{graphicx}
\usepackage{natbib}
\usepackage{doi}

\newlength{\figwidth}
\title{Measuring the spatial Acuity of vibrotactile Stimuli:\\ \emph{A new Approach to
determine universal and individual Thresholds}}

\newif\ifuniqueAffiliation
\uniqueAffiliationtrue

\ifuniqueAffiliation 
\author{
  \href{https://orcid.org/0000-0003-0687-0667}{\includegraphics[scale=0.06]{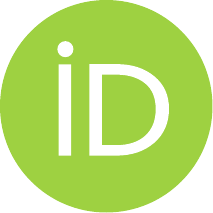}\hspace{1mm}
  Max vom~Stein}\thanks{Corresponding author: \texttt{mvomstein@uni-wuppertal.de}} \\
  Chair of Mechatronics\\
  University of Wuppertal
  \And
  Maximilian Hoppe\\
  Chair of Mechatronics\\
  University of Wuppertal\\
  \And
  Maxim Sommer\\
  Chair of Mechatronics\\
  University of Wuppertal\\
  \And
  Kai-Dietrich Wolf\\
  Chair of Mechatronics\\
  Institute for Security Systems\\
  University of Wuppertal
}

\else
\usepackage{authblk}

\newbox{\orcid}\sbox{\orcid}{\includegraphics[scale=0.06]{orcid.pdf}} 
\author[1,2]{%
	\href{https://orcid.org/0000-0003-0687-0667}{\includegraphics[scale=0.06]{orcid.pdf}\hspace{1mm}
	Max vom~Stein}\thanks{\texttt{mvomstein@uni-wuppertal.de}}}
\author[1]{Maximilian Hoppe}
\author[1]{Maxim Sommer}
\author[1,2]{Kai-Dietrich Wolf}

\affil[1]{Chair of Mechatronics, University of Wuppertal, Gaußstraße 20, Wuppertal 45131, Germany}
\affil[2]{Institute for security systems, University of Wuppertal, Talstraße 71, Velbert 42551, Germany}
\fi

\renewcommand{\shorttitle}{Measuring the spatial Acuity of vibrotactile Stimuli}

\hypersetup{
	pdftitle={Measuring the spatial Acuity of vibrotactile Stimuli: A
		new Approach to determine universal and individual Thresholds},
	pdfsubject={cs.HC},
	pdfauthor={Max vom~Stein, Maximilian Hoppe, Maxim Sommer,
		Kai-Dietrich Wolf}, 
	pdfkeywords={Tactile perception of vibration, Sensory
		testing, Vibrotactile spatial acuity, Tactile interface design, Human-machine
		interaction, Tactile displays, Sensory substitution, Bayesian adaptive parameter
		estimation},
		}

\begin{document}
\maketitle

\begin{abstract}
Tactile perception is an increasingly popular gateway in human-machine
interaction, yet universal design guidelines for tactile displays are still
lacking, largely due to the absence of methods to measure sensibility across
skin areas. In this study, we address this gap by developing and evaluating two
fully automated vibrotactile tasks that require subjects to discriminate the
position of vibrotactile stimuli using a two-interval forced-choice procedure
(2IFC). Of the two methodologies, one was initially validated through a
preliminary study involving 13 participants. Subsequently, we applied the
validated and improved vibrotactile testing procedure to a larger sample of 23
participants, enabling a direct and valid comparison with static perception. Our
findings reveal a significantly finer spatial acuity for static stimuli
perception compared to vibrotactile stimuli perception from a stimulus
separation of $15~mm$ onwards. This study introduces a novel method for
generating both universal thresholds and individual person-specific data for
vibratory perception, marking a critical step towards the development of
functional vibrotactile displays. The results underline the need for further
research in this area and provide a foundation for the development of universal
design guidelines for tactile displays.
\end{abstract}


\section{Introduction}
\label{sec1}
The human skin, as the body's largest and most accessible sensory organ,
presents unique challenges for researchers and professionals in tactile
perception and in human-machine interaction. Unlike vision and hearing,
standardized test procedures for measuring tactile perception are limited
(\cite{Wee.2021}). Nevertheless, ongoing research has uncovered insights that hold
potential for numerous applications in user-environment communication. The
haptic sense could complement or even replace other human senses, greatly
benefiting individuals with limitations in other sensory modalities
(\cite{Lederman.2009, Israr.2012,Johnson.2006,Filgueiras.2016}).

Various studies have already explored the use of tactile perception for
information transfer. For example, \cite{01_Adame.2013} designed a wearable belt
with vibrotactile motors, allowing subjects to successfully navigate a virtual
environment using only vibrations. \cite{12_Meers.2005} developed a substitute
vision system employing a stereo-camera's depth view to detect object distances
in images, transmitting this information to the wearer's fingertips via
electrical signals. Additionally, \cite{13_Pamungkas.2015} focused on enabling
tactile sensation of surface texture in the context of prosthetics and immersive
technology, devising a feedback system consisting of an artificial finger that
converts vibratory input signals into electrical pulses for wearer stimulation.

Despite these encouraging use cases of the human skin as a gateway for
transmitting information, it is crucial to ground practical applications with
physiological and neurological research on the human body. Although there
already exist several functional devices, targeted measurements need to be
carried out in order to optimize the tactile information transfer and minimize
the misrecognition rate by the user. Of particular interest are dermal tactile
perceptiveness measures, which provide information on perceptual limits and
possibilities. One approach to address this issue is to fit psychometric
functions of participants with discrimination tasks (cf. Ch.~\ref{sec2.2}). The
Two-Point-Discrimination (2PD) (\cite{04_Dellon.1978}) is a widely used
procedure for establishing benchmarks of tactile spatial acuity of static
(pressure) stimuli, requiring participants to differentiate between the
placements of one or two tips at varying distances on a selected skin site.
\cite{15_Tong.2013} identified significant weaknesses in the 2PD method and
proposed an alternative, the 2-point-orientation-discrimination (2POD), where
participants differentiate between the proximal-distal (hereafter referred to as
vertical) and medial-lateral (hereafter referred to as horizontal) placement of
two tips. By comparing the two methods, they demonstrated the 2PD's
susceptibility to non-spatial cues at minimal tip distances on the skin.
Furthermore, they utilized Kontsevich and Tyler's (\cite{10_Kontsevich.1999})
Bayesian adaptive testing, an algorithm that selects only the most informative
intensities or separations, thus minimizing trial costs and duration.

In this paper, two types of stimuli are distinguished: static and vibratory
stimuli. Static stimuli, characterized by their very low or non-existent
oscillation frequency ($<1~Hz$), describe the contact of a stationary object with
the skin. For instance, in Tong et al.'s experiments, such stimuli are produced by
pressing calipers onto the skin. On the other hand, vibrotactile stimuli are
created by objects in contact with the skin, exhibiting higher oscillation
frequencies ($>10~Hz$, in our case $\sim{}130~Hz$). The transducers (stimulus tips)
maintain consistent contact with the skin from the onset to the cessation of the
vibration stimulus (\cite{Schmidt.2007}).

Perception of static stimuli has been extensively researched across various skin
areas (\cite{03_Cholewiak.2003, 07_Johannesson.2017, 02_Cholewiak.2004}). However,
the practical application of static stimuli in tactile interfaces proves
somewhat challenging to implement. The actuation required for static stimuli is
bulky and is limited to conveying quasi-binary information. In contrast,
vibrotactile stimuli offer a more feasible solution. They allow additional
dimensions of information transfer through frequency, pulse patterns, and
intensity. Furthermore, vibration motors are easily controllable, compact in
size, and therefore integrate well into textiles or similar materials
(\cite{Kern.2023}).

Vibrations in the range of $20-50~Hz$ are predominantly perceived by Meissner's
corpuscles, while those in the range of $100-200~Hz$ are mainly detected by
Pacinian corpuscles. Conversely, static pressure stimuli are primarily
registered by Merkel's discs and Ruffini corpuscles. These receptors, aside from
their distinct sensitivities and receptive fields, also exhibit varied
distribution densities across different skin areas (\cite{Schmidt.2007}).

Therefore, benchmarks for static stimuli, such as the 2PD and 2POD, cannot
readily be applied to vibrotactile stimuli. These require separate investigation.
Unlike static stimuli, a comprehensive comparable investigation of various skin
areas' response to vibrotactile stimuli is yet to be undertaken.

Significant research has been carried out in the field of vibrotactile stimuli.
\cite{03_Cholewiak.2003} studied the impact of frequency,
location on the body, proximity along the loci, and observer age on spatial
acuity by introducing an array of tactors on the forearm. They found that
stimulus position and separation along sites predominantly influenced spatial
acuity. \cite{07_Johannesson.2017} utilized eccentric
rotating mass (ERM) motors with in-plane vibration, requiring observers to judge
which neighboring tactors were active as motor distances decreased. They noted
that discrimination accuracy remained well above chance at their smallest motor
distance. \cite{02_Cholewiak.2004} also employed a belt
equipped with tactors around subjects' abdominal areas, systematically modifying
the number and position of tactors to determine the most suitable loci for
installing a tactile display.

Within the broader scope of our research efforts, a key goal is the development
of expanded design guidelines for vibrotactile interfaces. Our ambition is to
strike an optimal balance, maximizing information transfer while minimizing the
misrecognition rate. For this purpose, establishing a comprehensive
understanding of vibrotactile acuity is indispensable.

While the studies mentioned previously have yielded valuable insights for
specific skin areas, their methodologies are not easily adaptable to new
conditions or settings, and they do not allow a direct comparison of benchmarks
due to their divergent principles. These methods are not designed for
large-scale studies that span multiple skin areas. Direct comparison with static
tactile perception using current studies proves challenging. The comparability
of methods for recording vibrotactile and static benchmarks is low. The
development of universal design guidelines for tactile interfaces is hindered by
the absence of methods for measuring sensibility across skin areas in multiple
directions.

\subsection{Objectives and Hypothesis}
\label{sec1.1}

Our research aims to address two key objectives, organized in a sequential
manner:

\begin{enumerate}
\item \textbf{Method Validation Experiment (MVE):} Our first goal is to develop
and validate a method that meets the specific needs of our investigation. This
method should embody the following features: 
  \begin{itemize}
  \item It can be applied to various skin areas
  \item It yields results that can be directly compared with studies examining
  the perception of static stimuli.
  \item It is capable of quantifying anisotropy effects.
  \item It is fully automated to minimize experimenter bias.
  \end{itemize}
Having designed multiple apparatuses, we aim to validate one through a dedicated
experiment. This experiment is designed specifically to evaluate the performance
of our proposed methods under controlled conditions. 

\item \textbf{Extended Perception Comparison Experiment (EPCE):} With the
validated method in place, our second goal is to apply this method to
investigate our primary research question. This involves conducting a
comprehensive experiment using our validated method to gather necessary data.
\end{enumerate}

Regarding our second objective, we have formulated the following hypothesis:

\vspace{1em}
\noindent\textbf{$H$}: \textit{The spatial acuity of human skin for vibrotactile
stimuli is significantly different from that for static stimuli.}
\vspace{1em}

This hypothesis embodies our theoretical expectation, rooted in the human
physiology concepts discussed earlier, and is the target of our empirical
testing using the method we plan to develop and validate as our first objective. 

To tackle our main objectives, this paper is structured into two consecutive sections:

\textbf{MVE}: To address the absence of suitable methods, we introduce two
methods applicable to various body locations that yield comparable results for
tactile acuity values. We have adapted Tong et al.'s (\cite{15_Tong.2013}) and
Dellon's (\cite{Dellon.1978}) principle of static stimulus presentation and
developed vibratory counterparts. Obtaining vibratory perception data requires a
significantly more complex apparatus setup. It is essential to avoid effects
from static perception, ensuring that detection is based solely on vibratory
sensations. We also employ Bayesian adaptive testing for selecting varying
distances between vibratory tips, which allows for more representative
measurements across different body locations in fewer trials
(\cite{10_Kontsevich.1999}). We designed the apparatuses with the intention to
support extensive studies. Their goals include acquiring vibratory acuity values
from various body areas, investigating perception anisotropy, and allowing
comparisons with electro-tactile and static perception. The apparatuses can be
converted to electro-tactile measuring with minimal effort.

\textbf{EPCE}: For this part, we adapt the validated vibrotactile
Two-Point-Discrimination (VT-2PD) (cf. Ch.~\ref{sec5.1.1}) method to match
Tong et al.'s conditions, allowing us to make a direct comparison with static
perception.


\section{Principles}
\label{sec2}

\subsection{Bayesian Adaptive Parameter Estimation (BAPE)}
\label{sec2.1}
We determined the psychometric functions of the participants by employing
Bayesian Adaptive Parameter Estimation (BAPE) with entropy minimization. This
procedure is based on the publication by \cite{10_Kontsevich.1999} and has been
successfully applied to tactile perception by \cite{15_Tong.2013}. We defined a
family of Weibull functions with the following form as the solution set for the
sought psychometric function:

\begin{equation}
  \psi_{a, b, \gamma}(x) = \gamma + (1 - \delta - \gamma)\left(1 - 2^{-\left(\frac{x}{a}\right)^b}\right)
  \end{equation}

\begin{figure}[!t] 
  \centering 
  \includegraphics[width=\figwidth]{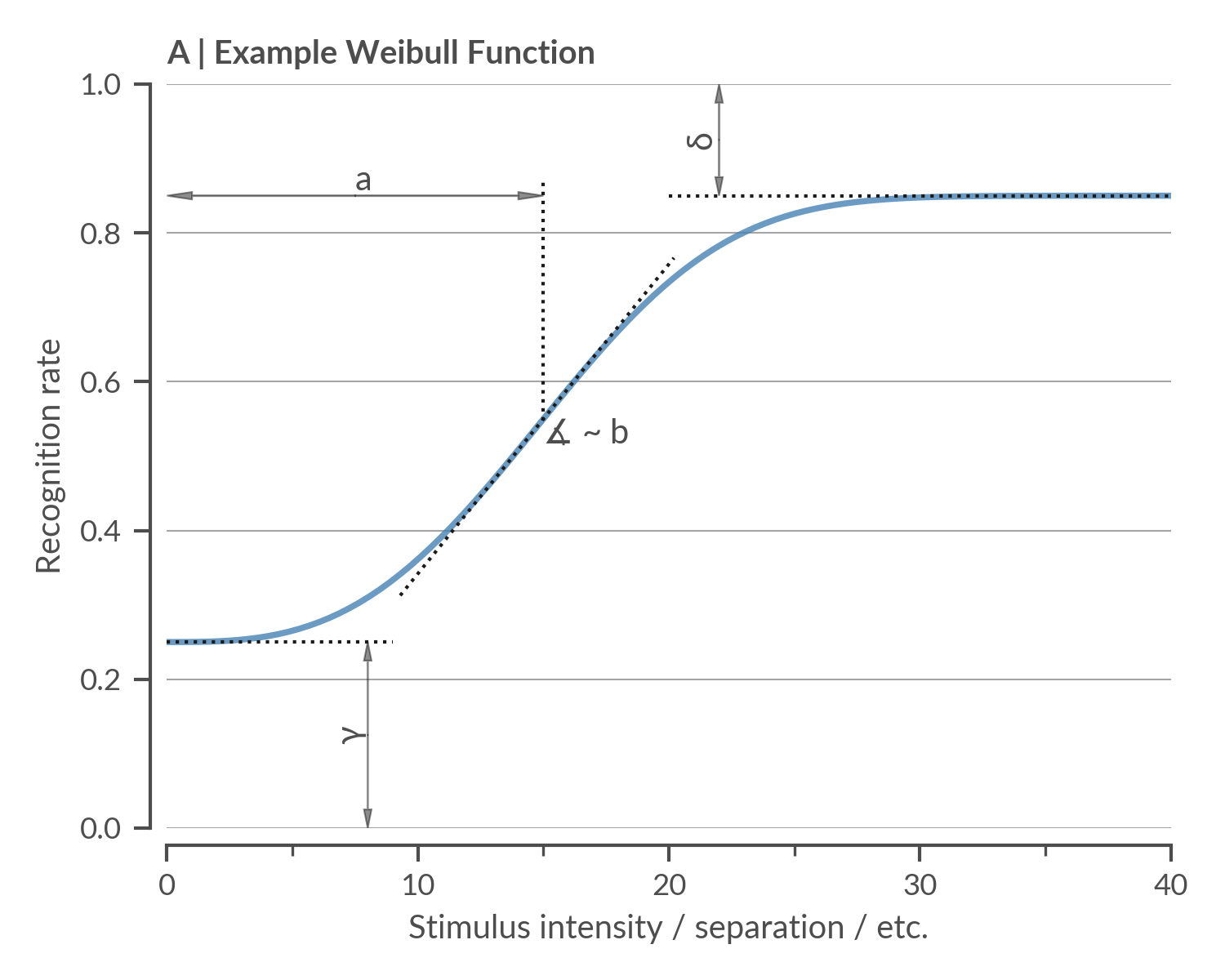}
  \caption{Example Weibull function as a possible psychometric function. $(\gamma)$~Guess rate. 
  $(\delta)$~Lapse rate. $(a)$~Threshold. $(b)$~Slope.}
  \label{fig_example_weibull}
  \end{figure}

The Weibull function is well-established as a psychometric function
(\cite{17_Wichmann.2001}) as all variable parameters have a directly discernible
influence on the curve's shape. An example function of the family is shown in
Figure~\ref{fig_example_weibull}, illustrating the properties of the individual function
parameters. The parameter $\gamma$ represents the guess rate, indicating the
detection rate at which the participant identifies the stimulus at lower
intensities, read off the function as the lower asymptote (e.g., $\gamma \approx
0.25$). It is opposed to $(1 - \delta)$, also called the lapse rate or finger
error, which indicates the recognition rate at the highest tested intensity and
thus maps the upper asymptote (e.g., $(1 - \delta) \approx 0.85$). The parameter
$a$ shifts the curve's inflection point in the x-direction and is often given as
a threshold value (e.g., $a \approx 15$). The parameter $b$ is related to the
slope in the transition region and stretches or compresses the curve. Using the
intervals and step sizes provided below, we created a family of curves with all
possible parameter combinations. We fixed $\delta$ at a constant value,
resulting in 90,000 psychometric functions. The priori distribution $P(\psi_{a,
b, \gamma})$ of the psychometric functions was specified as uniformly
distributed.

In the test procedure, one response from the participant was required per test
query, which was scored as correct or incorrect. After each query, we calculated
the probability space $p_t(a, b, \gamma)$, which indicates the probability of
occurrence of each parameter combination and, consequently, the psychometric
function based on the participant's response. We evaluated the entropy $H_t$ by
using $p_t(a, b, \gamma)$ (\cite{brand.1999}):

\begin{equation}
  H_t = -\sum_{a, b, \gamma} p_t(a, b, \gamma) \log(p_t(a, b, \gamma))
  \end{equation}

In this case, smaller entropies indicate higher agreement. The entropy change
between two trials indicates the information gain. We determined the subsequent
query by minimizing entropy, thus maximizing the information gain. Based on the
responses, we determined the posteriori probabilities of the psychometric
functions $P(\psi_{a, b, \gamma} | r_i)$ and expected probabilities for the
function parameters $a$, $b$, and $\gamma$. Proportionally to the probabilities
of all possible Weibull functions, we formed the so-called postmean from the set
of curves. Thus, we assigned a specific psychometric function to each
participant.

The participants underwent 50 trials in the two-interval forced-choice (2IFC)
procedure (\cite{14_Ratcliff.2018}). We determined the distances to be tested
($d_{2PD}$ and $d_{2POD}$, see Fig~\ref{fig_principle_vt-2pd_vt-2pod}) in both
procedures using the BAPE algorithm. The interval limits for the guess rate
$\gamma$ were set from $0.01$ to $0.99$, divided into 100 steps. We fixed the
finger error $\delta$ at $0.02$ based on recommendations by
\cite{09_Klein.2001}. We set the limits for the threshold $a$ at $2.5$ to
$45~mm$ in 18 steps and for the slope $b$ at $0.01$ to $10$ in 50 steps.

\subsection{Experimental Designs}
\label{sec2.2}

\subsubsection{Experimental Design MVE}
\label{sec2.2.1}


\begin{figure}[!t]
  \centering
  
  \begin{minipage}{0.49\textwidth}
    \includegraphics[width=\linewidth]{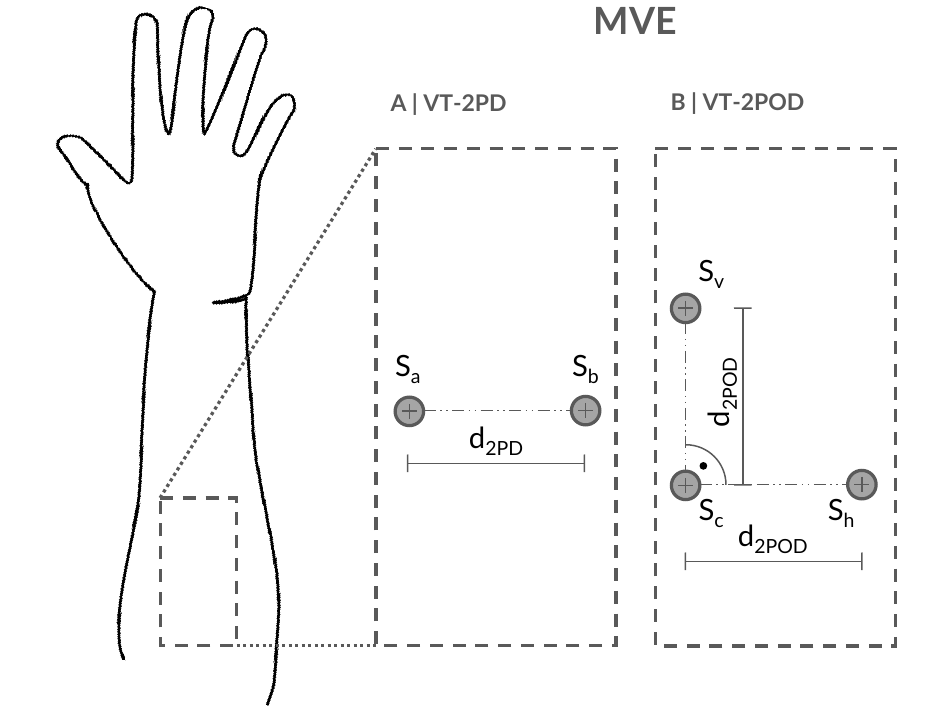}
    \caption{Schematic representation of the experimental tasks \mbox{VT-2PD} and
    \mbox{VT-2POD} for the Method Validation Experiment (MVE). Assessment conducted on the
    left forearm. Placement of the stimulus tips for \mbox{VT-2PD} (\textbf{A}) and
    \mbox{VT-2POD} (\textbf{B}) illustrated.}
    \label{fig_principle_vt-2pd_vt-2pod}
  \end{minipage}
  \hfill 
  \begin{minipage}{0.49\textwidth}
    \includegraphics[width=\linewidth]{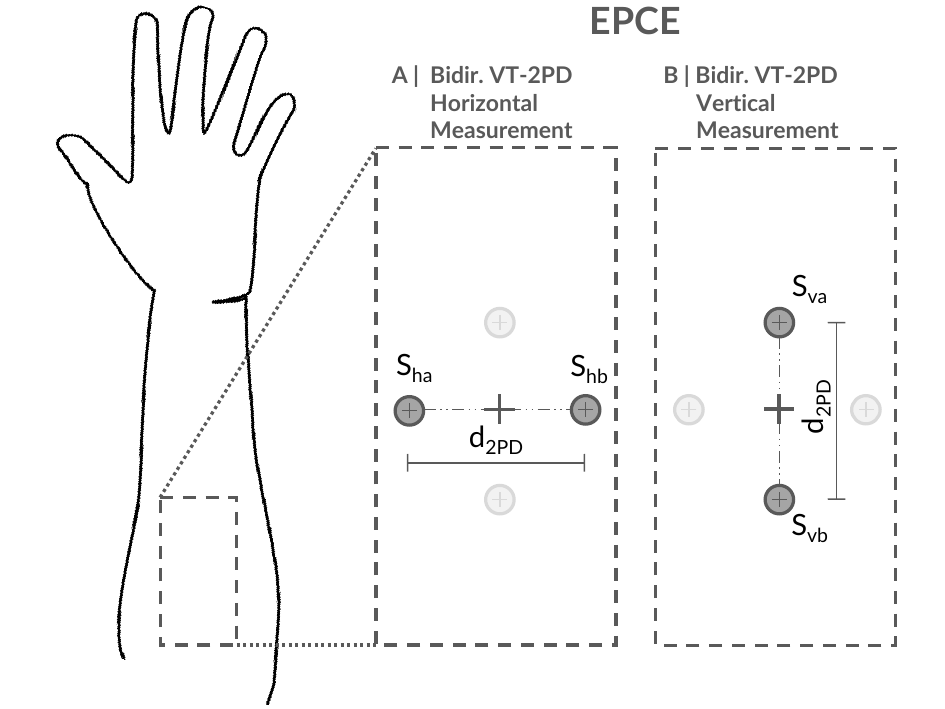}
    \caption{Schematic representation of the experimental task bidirectional
    \mbox{VT-2PD} for the Extended Perception Comparison Experiment (EPCE).
    Assesment conducted on the left forearm. Consecutive placement of the stimulus
    tips as (\textbf{A}) and (\textbf{B}). Order is randomized in the experiment.}
    \label{fig_principle_bidir_vt-2pd}
  \end{minipage}
  
\end{figure}
  
In their study, \cite{15_Tong.2013} compared the traditional clinical method of
Two-Point-Discrimination (2PD) with their novel method,
Two-Point-Orientation-Discrimination (2POD), recording threshold values for
tactile perception sensitivity to static stimuli. To establish direct
comparability, our experiment for determining threshold values for vibratory
stimuli is based on Tong et al.'s work. We focused on the skin area of the left
forearm and adapted our first method from the 2PD task (\cite{04_Dellon.1978}).
As illustrated in Figure~\ref{fig_principle_vt-2pd_vt-2pod} (\textbf{A}), our
design employs two stimulus generators, $S_a$ and $S_b$, placed on the skin at
varying distances ($d_{2PD}$). For each specified distance, a query is conducted
by randomly emitting a vibration pulse from first one generator, then the other
(2IFC)(\cite{Schlauch.1990}). The participant must identify which side was
stimulated first. We named this method vibrotactile Two-Point-Discrimination
(VT-2PD). It is important to note that, unlike the static 2PD model procedure,
there is no intensity bias in this method. In each query, both sides are
activated, with only the order being randomized. Both stimulus tips have full
contact pressure while the query is executed.

We based the principle of the second experiment on the 2POD task
(\cite{15_Tong.2013}). As depicted in Figure~\ref{fig_principle_vt-2pd_vt-2pod} (\textbf{B}),
three signal generators are arranged in an isosceles triangle with adjustable
varying distances ($d_{2POD}$). Participants must distinguish between horizontal
and vertical stimulus inputs. This is achieved by activating signal generator
$S_c$ in each stimulus pair along with either randomly selected signal generator
$S_v$ or signal generator $S_h$. We named this method vibrotactile
Two-Point-Orientation-Discrimination (VT-2POD). Here also all three stimulus
tips have full contact pressure while the query is executed.

\subsubsection{Experimental Design EPCE}
\label{sec2.2.2}

Upon the successful validation of our vibrotactile Two-Point-Discrimination
(VT-2PD) method, as detailed in chapter~(cf. Ch.~\ref{sec5.1.1}), our goal was to juxtapose our
findings of vibrotactile sensibility with those of Tong et al.'s work on the forearm's
sensibility to static stimuli. To facilitate this comparison in the Extended
Perception Comparison Experiment (EPCE), we took measures to ensure the
compatibility of our method with Tong et al.'s Two-Point-Orientation-Discrimination
(2POD) procedure.

Tong et al.'s approach involved discerning between horizontal and vertical
stimuli, a process that inherently neutralized potential anisotropic effects. On
the other hand, our VT-2PD task was constructed with anisotropy in mind,
examining only one directional axis on the skin. To ensure a meaningful
comparison with Tong et al.'s 2POD task, we expanded our procedure to probe two
axes on the skin, subsequently offsetting anisotropic effects by generating the
psychometric functions including both directions.

While the fundamental design of the VT-2PD remained consistent with that
outlined in Chapter~\ref{sec2.2.1}, our approach involved the consecutive
measurement in two directions. The stimuli were introduced sequentially in two
orientations - (\textbf{A}) horizontal and (\textbf{B}) vertical
(see Fig.~\ref{fig_principle_bidir_vt-2pd}). The sequence of these orientations was
randomized for each participant. Both orientations shared the same center point,
creating a '+'-formation on the skin. This approach allowed us to retain the
primary structure of the VT-2PD method while accommodating the requirements for
a fair comparison with Tong et al.'s 2POD task.

\section{Material and Methods}
\label{sec3}
This chapter outlines the materials and methodologies used in both our studies
MVE and EPCE. We detail the apparatus, experimental procedures, and participant
selection, providing a basis for understanding our results and findings.

\subsection{Apparatus}
\label{sec3.1}
The technical setup consisted of two main elements: the calipers, which guide
and position the stimulus generators, and the lifter, which moves the calipers
vertically and establishes a consistent contact pressure of the stimulus tips on
the subject's skin.

\begin{figure*}[!t] \centering
  \includegraphics[width=\textwidth]{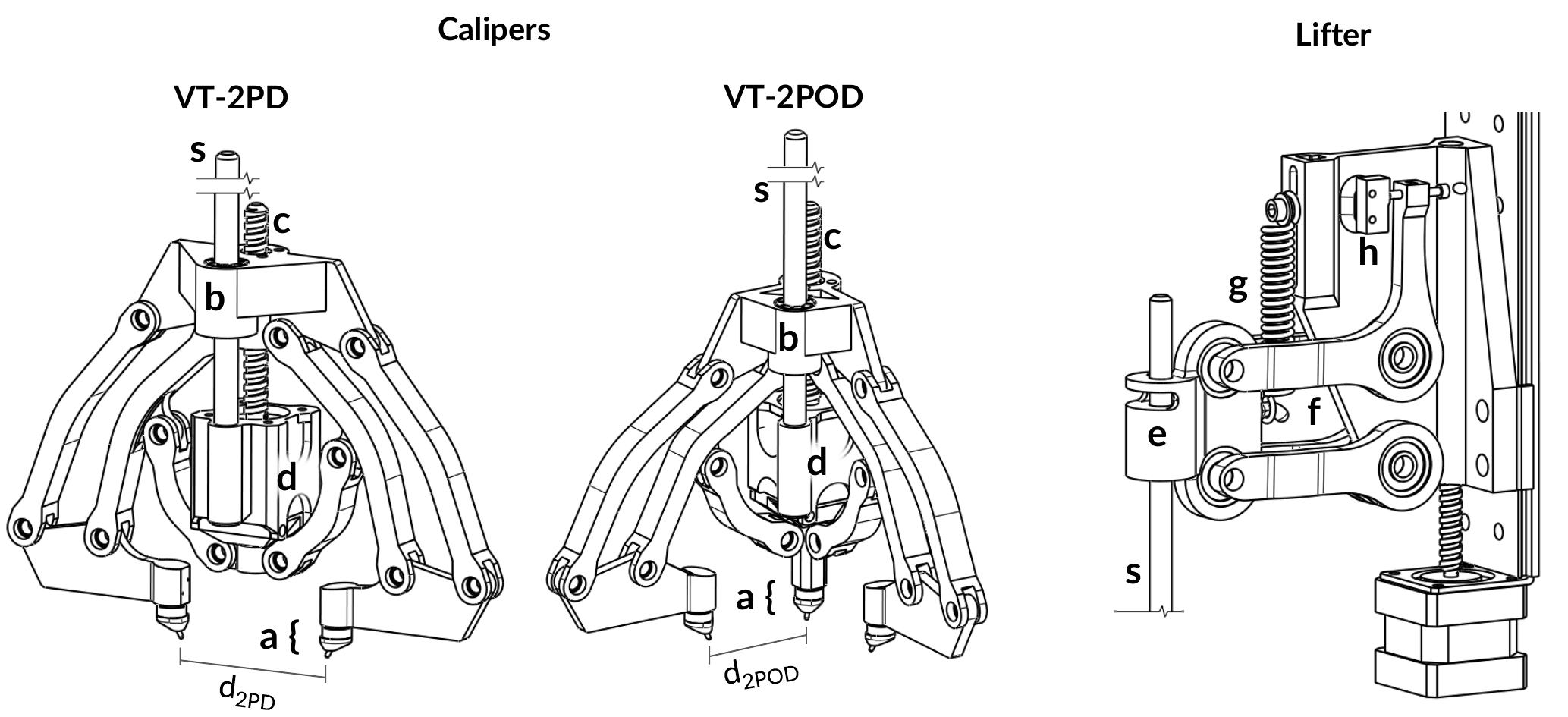} \caption{Calipers for
  VT-2PD and VT-2POD trial. Both mounted with axis \textbf{(s)} to the lifter in
  order to be able to move in height. \textbf{(a)} Stimulus generator.
  \textbf{(b)} Linear guide. \textbf{(c)} Spindle drive for distance adjustment.
  \textbf{(d)} Stepper motor. \textbf{(e)} Caliper attachment. \textbf{(f)}
  Four-bar linkage. \textbf{(g)} Tension spring for calibration of contact
  pressure. \textbf{(h)} Pushbutton for detecting the contact pressure.}
  \label{fig_tech_draw_apparatus}
  \end{figure*}

\subsubsection{Calipers}
\label{sec3.1.1}
In both the 2PD and the 2POD configurations, two six-bar linkages guide two
stimulus generators (\textbf{a}; see Fig.~\ref{fig_tech_draw_apparatus}), each
consisting of a stimulus tip attached to a vibration motor. They can be adjusted
in a common plane without rotational motion or tilting. The head of the coupling
gear is linearly guided (\textbf{b}) and positioned via a spindle drive
(\textbf{c}) with a stepper motor (\textbf{d}) located in the center.

\begin{figure}[!t] 
  \centering
  \includegraphics[width=\figwidth]{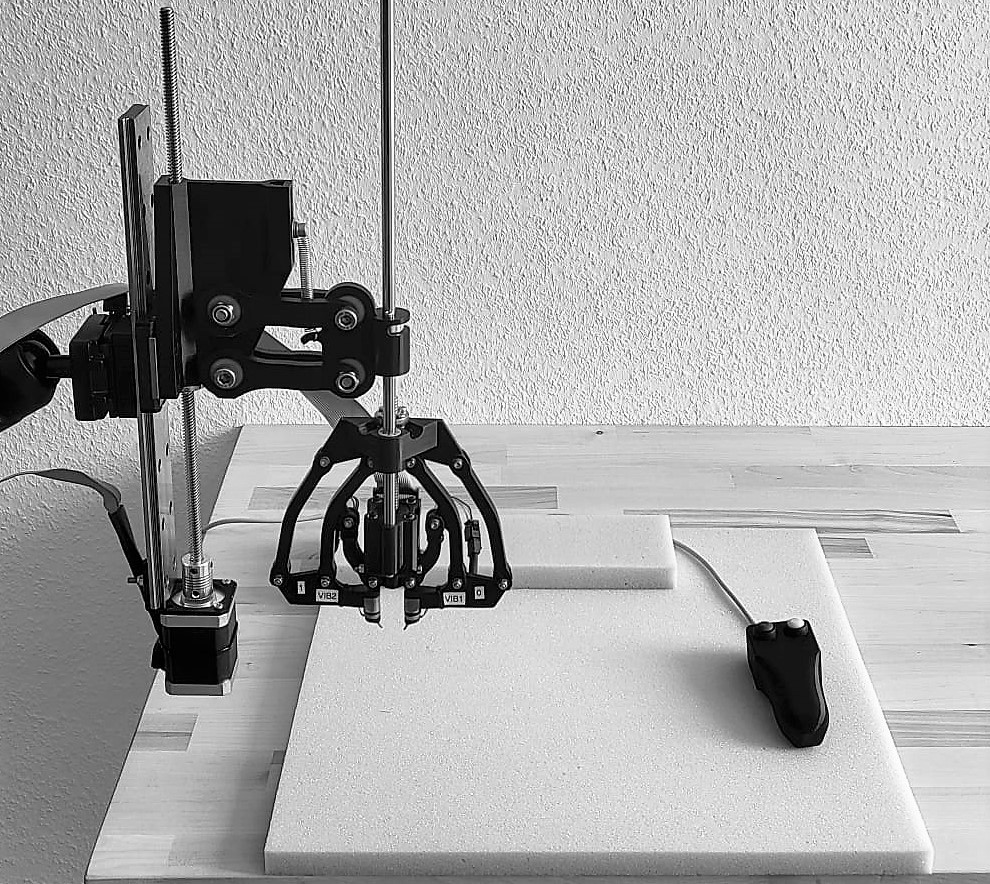} \caption{Full
  experimental setup, featuring the VT-2PD calipers currently installed.
  Participant's response remote is located on the right.} 
  \label{fig_setup_apparatus}
  \end{figure}

\subsubsection{Lifter}
\label{sec3.1.2}
The lifter allows the caliper to move linearly in height. The guide rail is
attached to a stand with a ball joint (see Fig.~\ref{fig_setup_apparatus}),
enabling flexible alignment of the entire apparatus. The caliper's clamping
(\textbf{s} in \textbf{e}, see Fig.~\ref{fig_tech_draw_apparatus}) is supported
by a four-bar linkage (\textbf{f}) and hangs with its self-weight on an
adjustable tension spring (\textbf{g}). This suspension permits minimal up and
down movements with very low breakaway torque. The spring preload can be used to
calibrate the contact pressure of the stimulus tips on the skin area. When the
desired contact pressure is reached during lowering, the pushbutton (\textbf{h})
is triggered, and the spindle drive stops.

\subsubsection{Potential for Adaptation to Electrostimulation}
\label{sec3.1.3}
In addition to its current configuration for vibrotactile experiments, the
design of our apparatuses allows for easy adaptation for future research
involving electrostimulation. With minimal modifications, the stimulus
generators could be replaced or supplemented with electrodes, expanding the
range of sensory modalities that can be investigated with our method.

\subsubsection{Components and Values}
\label{sec3.1.4}
The stimulus is applied using eccentric rotating mass (ERM) vibrators with a
diameter of $10~mm$ and a weight of $0.9~g$ (\cite{Stronks.2015}). We operate
these at a voltage of $5~V$ and adjust the RPM using pulse width modulation
(PWM). The average frequency with no load is $131~Hz$. A NEMA 11 stepper motor
is used in the caliper, and a NEMA 17 stepper motor in the lifter. All flexible
pivot points are double-sided with radial ceramic bearings to ensure the lowest
possible friction and precise guidance. The ERM vibrators are connected to the
guide with a damper to decouple vibration. Stimulation is transmitted from the
ERM vibrator to the skin through a rounded spring steel tip of $1.5~mm$ in
diameter. The stimulus tips' center points can be moved in an interval of
$2.5~mm$ to $60~mm$ with a tolerance of $< 0.2~mm$. We set the contact pressure
to $0.5~N$ via the tension spring's position, which is achieved with a tolerance
of $< 4~\%$. This ensures sufficient contact pressure of the stimulus tips and
does not cause discomfort for the participants. The length of the vibration
pulses is set to $200~ms$. We use an NVIDIA Jetson Nano developer board to drive
the components, providing sufficient processing capacity to perform the more
computationally intensive entropy minimization in a reasonable amount of time.
We programmed the application exclusively in Python 3 (\cite{Python.2009}). All
structural elements of the experimental apparatus are 3D printed using fused
deposition modeling (FDM). Furthermore, only off-the-shelf components were used
to allow other researchers to quickly reproduce the apparatus.


\subsection{Participants and Procedure}
\label{sec3.2}

\subsubsection{Participants and Procedure MVE}
\label{sec3.2.1}
To evaluate our test procedures, a preliminary study was conducted on 13 
participants, consisting of 8 men and 5 women aged 21 to 33, with an average age
of 28 years. All participants were right-handed and had no medical history of
restrictive diseases affecting tactile perception (e.g., diabetes or carpal
tunnel syndrome) or cognitive perception disorders (central nervous system
disorders, learning disabilities, attention deficit disorder, or dyslexia)
(\cite{05_Grant.1999}). The tested skin areas were free of skin irritation,
tattoos, and scar tissue. All subjects were instructed and signed informed
consent and privacy statements.

After the participant had completed the self-report form, she*he was seated and
her*his left forearm was comfortably placed on a foam pad on the table with the
palm facing up. The experimental apparatus was aligned perpendicularly on the
skin area on the forearm, ensuring uniform contact pressure of the stimulus tips
at all distances. The participant was given the remote control for response input
in the right hand and performed a first run with ten trials to understand the
experiment and was then provided with noise-canceling headphones playing white
noise. The experimenter then started the experiment. 

We implemented the following steps fully automated in the program code to
eliminate bias or influence by the experimenter. The BAPE algorithm determines
the distance to be selected for the stimulus tips with the highest information
content. This is set, and then the calipers are lowered onto the subject until
the desired contact pressure is achieved. For VT-2PD, a random burst is first
delivered on one side, and then on the other side (cf. Ch~\ref{sec2.2}). For
VT-2POD, a random burst orientation is chosen. The participant enters their
response via the remote control, and the lifter raises the calipers again. The
BAPE algorithm determines the new distance based on the user response, and the
loop runs again until the required number of trials is reached. Then the
determined postmean function is returned (cf. Ch~\ref{sec2.1}), and the
data is saved under an anonymous participant identification (test subject
identification, TSID).

\subsubsection{Participants and Procedure EPCE}
\label{sec3.2.2}
In the Extended Perception Comparison Experiments (EPCE), we engaged a group of
23 participants (18 males, 5 females) aged between 18 and 38 years to determine
the quantitatively meaningful vibrotactile perception acuity. The participants
were financially compensated for their time commitment. It is important to note
that none of the participants from the prestudy were re-examined in this phase
of the study.

To compare the results of our validated method with those of static stimuli, we
conducted the VT-2PD experiment again in a bidirectional form. The procedure was
generally the same as in the MVE. The orientation axis (horizontal or vertical)
was randomly determined at the beginning, and trials were conducted. After 50
queries, the participant remained seated while the VT-2PD apparatus was
reoriented. The apparatus was rotated by 90 degrees, and another 50 queries were
conducted.

The orientation of the buttons on the remote control could also be rotated
accordingly to avoid confusion. The psychometric function was then formed from
all 100 queries, allowing for a direct comparison with Tong et al.'s method. 

\subsection{Preventive Steps for valid Results}
\label{sec3.3} 

\subsubsection{Correct below-Threshold Detection}
\label{sec3.3.1} 
In both experiments, we need to control two main sources of correct
below-threshold detection: 1) the subject identifying the stimulus with the help
of other senses, and 2) the subject recognizing one-sided tactile motor
characteristics. Other senses can be easily blocked by using a blindfold and
noise-canceling headphones playing white noise. Even precisely manufactured
vibration motors have deviating characteristics, especially in amplitude and
frequency. We have examined identical motors in advance and selected those with
the highest agreement. Driving the motors via PWM allows us to calibrate their
amplitude and frequency. To further refine the control and prevent participants
from learning specific motor characteristics, we introduce a random variation to
the intensity of the motors. This variation is based on a Gaussian distribution,
with the duty cycle ratio serving as the expected value and a standard deviation
of 3. By selecting a value within this distribution for each trial, we ensure a
slight random fluctuation in motor intensity, effectively masking the motor
characteristics and mitigating learning effects.

\subsubsection{One-sided Application Pressure Bias}
\label{sec3.3.2}
In the application of the VT-2PD methodology, it is crucial to mitigate the risk
of one-sided bias that could compromise the validity of the results. To this
end, we have implemented several precautionary measures, starting with the EPCE,
in addition to employing the 2IFC procedure, which is inherently designed to
reduce bias \cite{Macmillan.2004}.

\paragraph{\textbf{Alignment Process}} A meticulous alignment process is
conducted to ensure uniform pressure application across different stimulus
separations. The experimenter initially aligns the apparatus manually on the
skin area, which is then verified with a short automated program. This program
sequentially moves through stimulus separations in five steps, from the largest
to the smallest, and lowers the stimulus tips onto the skin. Following the
methodology outlined in Tong et al. \cite{15_Tong.2013}, the experimenter
monitors the penetration depth of the tips to ensure consistent applied pressure
across all stimulus separations.

\paragraph{\textbf{Monitoring Parameters}} While we believe that the variations
in applied pressure are not as significant as those encountered in manual
application of stimuli, it remains important to acknowledge that achieving
perfectly uniform pressure cannot be fully guaranteed. To address this issue,
the experimenter carefully monitors three key parameters:

\begin{itemize}
\item Visual inspection of output data is conducted after each test run to
identify frequent one-sided biased answers, particularly at recurring stimulus
separations.
\item A binomial test ($\alpha = 0.05$) is conducted to determine if one side is
preferentially chosen in general \cite{howell.2012}.
\item Anomaly detection in response times, flagged by a t-test, serves as an
additional indicator for the experimenter \cite{howell.2012}.
\end{itemize}

\paragraph{\textbf{Exclusion Criteria}} Test runs exhibiting indications of bias
are systematically excluded from the final analysis to preserve the integrity of
the study.\\\

By implementing these preventive measures, we aim to enhance the robustness and
validity of our findings, thereby contributing to the reliability of the VT-2PD
methodology in tactile perception research.

\section{Results}
\label{sec4}

\subsection{Results MVE}
\label{sec4.1}
\subsubsection{Exemplary individual Test Runs}
\label{sec4.1.1}

\begin{figure*}[!t]
  \centering
  \includegraphics[width=\textwidth]{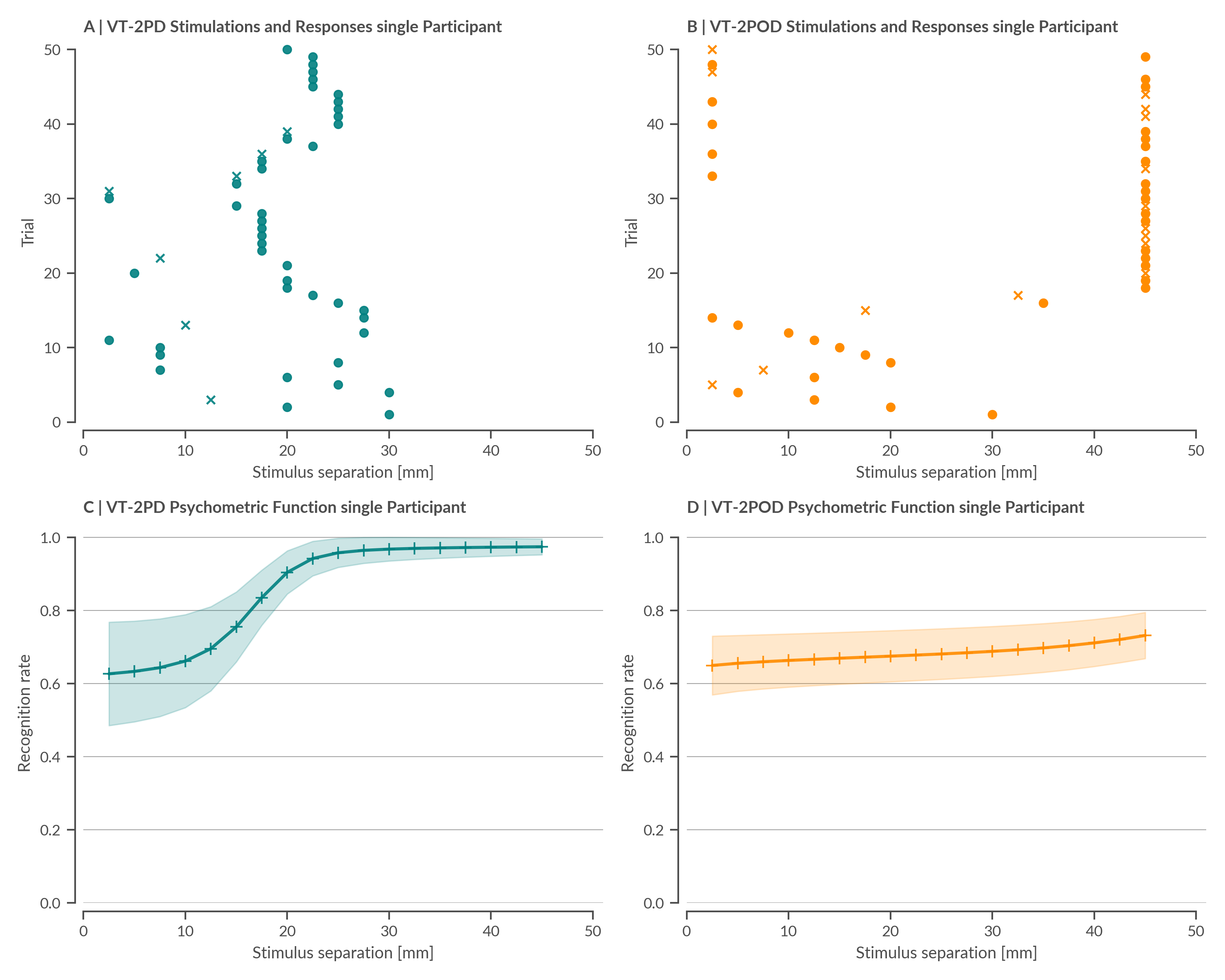}
  \caption{\textbf{MVE}: Exemplary diagrams of an individual VT-2PD experiment
  (\textbf{A}) and an individual VT-2POD experiment (\textbf{B}). Participant
  response at each respective trial is plotted over stimulus separation. Correct
  responses are marked with circles, while incorrect responses are marked with
  crosses. Point accumulation illustrates the search focus area of the BAPE
  algorithm. Lower plots (\textbf{C}, \textbf{D}) display the resulting
  psychometric function (\textbf{C}: teal, \textbf{D}: orange) with $\pm
  1$ SE (\textbf{C}: light teal, \textbf{D}: light orange) for individual
  participants' test runs. Plus symbols (+) denote the separations at which
  measurements were taken.}
  \label{fig_individual_vt2pd_vt2pod}
  \end{figure*}

Figure~\ref{fig_individual_vt2pd_vt2pod} (\textbf{A} and \textbf{B}) illustrates
the automated BAPE algorithm procedure based on the participants' responses. If
the participants' responses become increasingly accurate between iterations, the
algorithm tends to investigate smaller distances and vice versa. 

In one execution of the VT-2PD (\textbf{A}, \textbf{C}), the exemplary search
pattern is evident. The query stabilizes in the transition region of the
psychometric function (\textbf{C}), near the mean threshold. The participant's
guess rate stands at $0.62$, and at a $23~mm$ stimulus separation, the $0.95$
level of the recognition rate is surpassed. The standard deviation is
substantial for small stimulus separations and diminishes towards the top,
attributable to the low polling frequency of small separations. 

The search pattern in the displayed VT-2POD run (\textbf{B}, \textbf{D})
exhibits bipolar behavior, starting at Trial 18. The BAPE algorithm conducts
queries exclusively at the maximum ($45~mm$) and minimum ($2.5~mm$) of the query
interval. Consequently, the participant demonstrates a high error rate even at
the maximum distance. The guess rate is at $0.65$, and the recognition rate does
not exceed $0.75$ with increasing stimulus separation.

\begin{figure*}[!t]
  \centering
  \includegraphics[width=\textwidth]{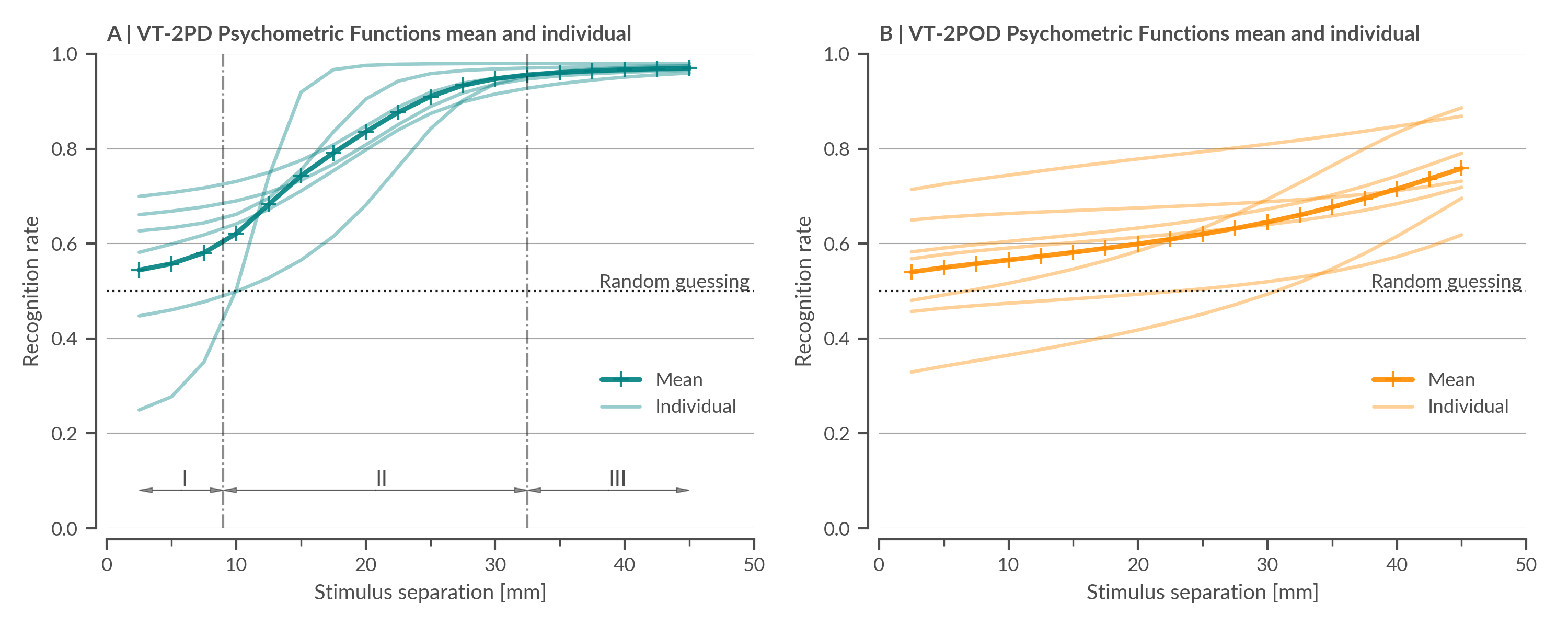}
  \caption{\textbf{MVE}: Plot of psychometric functions for individual participants and mean
  for (\textbf{A}) VT-2PD and (\textbf{B}) VT-2POD. The $0.5$ threshold (random
  guessing) of recognition rate is marked with a dotted line in both plots.
  Clear s-shaped psychometric functions are observed for VT-2PD but not for
  VT-2POD. The VT-2PD diagram is divided into three intervals, while no
  separation is possible for the VT-2POD diagram. Quantitative results are not
  significant for physiology due to the small number of participants. Plus
  symbols (+) denote the separations at which measurements were taken.}
  \label{fig_psyfun_vt2pd_vt2pod}
  \end{figure*}

\subsubsection{Individual and mean Psychometric Functions}
\label{sec4.1.2}
Figure~\ref{fig_psyfun_vt2pd_vt2pod} presents individual and mean psychometric
functions for both the VT-2PD and VT-2POD experiments, derived from the Weibull
function. Participant-specific psychometric functions were computed iteratively
using the BAPE algorithm across 50 queries during each experimental run. The
individual curves (light teal \textbf{A}, light orange \textbf{B}) were averaged
(teal \textbf{A}, orange \textbf{B}) in both trials.

For the VT-2PD experiment (\textbf{A}), both the individual participant curves
and the mean exhibit the common s-shape for psychometric functions (cf.
Ch.~\ref{sec2.1}). As such, we can divide the graph areas into three distinct
segments. Segment I represents the participants' sub-threshold level, falling
within the random guessing ($0.5$) region and indicating minimal or negligible
stimulus recognition. Segment II includes the transition region,
where the steepest slope signifies a rapid increase in recognition as distance
expands. In Segment III, participants exceed the $0.95$ recognition rate
threshold. When defining physiological processes, this threshold is frequently
employed as the recognition level (\cite{Crawford.2009}). Four individual
participant curves closely cluster together and approximate the mean, while two
curves diverge more significantly. The mean guess rate $\gamma$ (cf.
Ch.~\ref{sec2.1}) for the collected data is $0.54$. The mean value surpasses the
$0.95$ recognition rate level at $32.5~mm$.

In contrast, the VT-2POD test (\textbf{B}) displays neither discernible
psychometric functions in the mean nor for individual participants. As a result,
we cannot subdivide the graph areas as we did for the VT-2PD (\textbf{A}).
Although some stimulus detection occurs, as the curves at greater distances
notably exceed random guessing, no participant surpasses the $0.95$ recognition
rate level. The dispersion of the curves is considerably high, preventing any
direct correlation.

We exclusively employ the results for functional verification of the
experimental procedures. Owing to the small number of participants, quantitative
statements regarding physiological perception are not feasible at this stage
(\cite{Green.1966}). Furthermore, we have chosen not to report standard deviations
as they would not be representative given the limited participant pool. It is
important to note that these results were not included in the EPCE, primarily
due to enhancements in our methodology and to prevent a potential bias from
training effects. This decision aligns with our commitment to ensure the
robustness and validity of our findings.

\subsection{Results EPCE} 
\label{sec4.2}

\begin{figure}[!t]
  \centering
  \includegraphics[width=\figwidth]{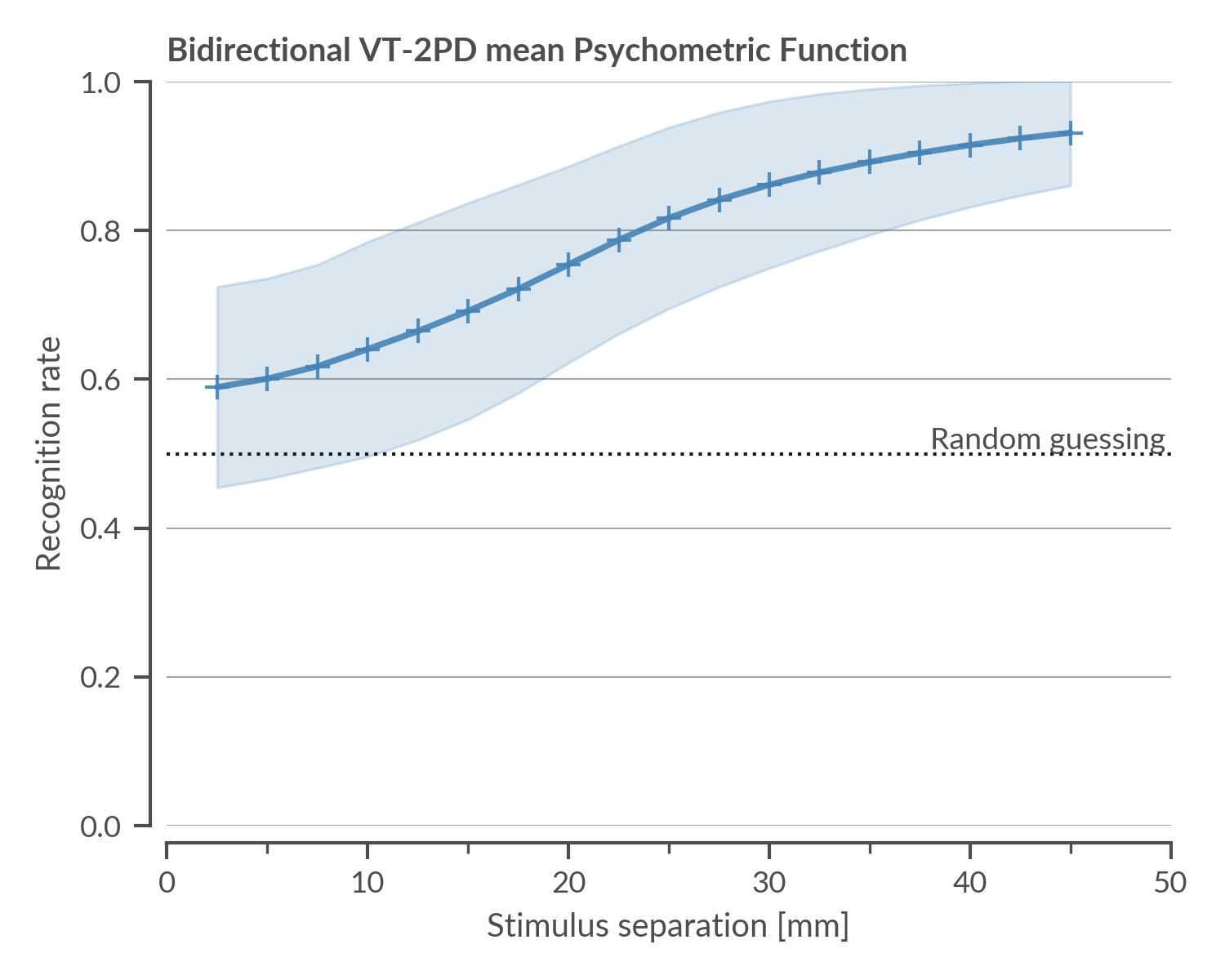}
  \caption{\textbf{EPCE}: Plot of mean psychometric function for bidirectionally
  (horizontally and vertically) applied VT-2PD. $\pm 1$ SE in light blue. The
  $0.5$ threshold (random guessing) of recognition rate is marked with a dotted
  line. Plus symbols (+) denote the separations at which measurements were
  taken.}
  \label{fig_psyfun_bidir_vt2pd}
  \end{figure}

Figure~\ref{fig_psyfun_bidir_vt2pd} presents the averaged psychometric function
of the bidirectional VT-2PD experiment. The guess rate starts at $0.59$ at a
distance of $2.5~mm$, and the maximum recognition rate escalates to $0.93$ at $45~mm$.
However, the $0.95$ threshold was not reached, potentially reflecting the impact
of the added complexity of vertical measurements.

The curve ascends notably more slowly than the curve from the purely horizontal
VT-2PD experiment (cf. Fig.~\ref{fig_psyfun_vt2pd_vt2pod}), indicating that the
additional vertical measurements may have impacted the detection sensitivity or
threshold. This divergence in gradient suggests a measurable discrepancy between
the horizontal and vertical measurements, emphasizing the challenge posed by
multi-directional vibrotactile perception.

Three distinct segments can be observed in the curve, as also shown in
Figure~\ref{fig_psyfun_vt2pd_vt2pod}, though they are not as clearly defined as
in the purely horizontal test condition. The transition region is noticeably
more pronounced, which could be interpreted as an indicator of the added
intricacy of combining vertical and horizontal measurements.

The standard deviation remains relatively high throughout the measurements,
which might be indicative of the inherent differences between horizontal and
vertical measurements. This high variability underscores the complexity of
accurately measuring and interpreting vibrotactile perception when multiple
directional stimuli are involved.

\section{Discussion}
\label{sec5}
\subsection{Discussion MVE}
\label{sec5.1}
\subsubsection{Quality of the Experiments}
\label{sec5.1.1}
Despite the relatively small sample size, the results of the Method Validation
Experiment (MVE) primarily support the conceptual validity of the VT-2PD
experimental setup over the VT-2POD.

\paragraph{\textbf{VT-2PD}}
\label{sec5.1.1.1}
The VT-2PD experiment demonstrates a clear tendency towards a $50~\%$ detection
rate below the perception level, which can be considered a goodness-of-fit
measure with an argument against strong bias. The measures against unintentional
correct below-threshold detection prove to be effective. Furthermore,
reproducible psychometric functions are obtained in the VT-2PD trial, clearly
displaying the three perceptual intervals: random, transitional, and detection
sections. Multiple experimental runs with the same participant exhibit minor
differences, supporting the experiment's reproducibility.

A weakness of the experiment lies in the precise alignment of the experimental
apparatus with the participant, as any misalignment can lead to a one-sided bias
caused by increased contact pressure. After several runs, the experimenter
became adept at alignment, and the error did not affect the results. False
responses from the participant at higher distances can cause significant
distortion of the psychometric function. One potential solution is to increase
the fixed delta value (cf. Ch.~\ref{sec2.1}) of the Weibull parameters,
thereby generating more tolerance for so-called finger errors.

The individual psychometric functions exhibit a high degree of variability in
the guess rate, leading to a substantial standard deviation in the derived mean.
This variability can be attributed to the limited number of queries conducted at
lower stimulus separations. To address this issue in the EPCE, we adjusted the
hyperparameters of the BAPE algorithm. This adjustment ensures more frequent
querying of the peripheral areas, thereby providing a more comprehensive
understanding of the perception thresholds across the entire range of stimulus
separations.

There is a known bias for the static 2PD experiment (\cite{15_Tong.2013}) due to
an intensity difference between the two possible queries in the task, which the
participant can recognize. However, this bias does not apply to the VT-2PD
experiment. Using the 2IFC method, both stimulus generators are activated in
every query, and both stimulus tips maintain constant contact with the skin,
ensuring no intensity bias arises.

\paragraph{\textbf{VT-2POD}}
\label{sec5.1.1.2}
In contrast, the VT-2POD experiment yields unreliable results. Participants
report that they do not perceive any horizontal or vertical orientation of the
stimulus, but only focus on the two outer stimulus generators. This results in
the principle of the VT-2POD trial being given with an increased minimum
distance by a factor of $\sqrt{2}$. Detection at greater distances exhibits
relatively low accuracy, and the improvement in detection is not substantial
compared to that at smaller stimulus separations. The psychometric functions are
not reproducible in the same participant and do not display the expected basic
s-shape.

The primary issue with the experiment is the uneven contact pressure on the
curved forearm surface. An alignment that ensures all three stimulus tips have
sufficient contact pressure at all distances of the interval is difficult or,
depending on the participant's physiology, impossible. At higher distances, a
stimulus tip has insufficient contact with the skin, leading to an inaccurate
detection rate. Trial runs of the VT-2PD with orientation vertical instead of
horizontal on the arm reveal significantly worsened recognition rates for
identical participants. This suggests an anisotropic perception of vibratory
stimuli in the examined skin area. If further studies confirm this finding,
there will be a directional bias, limiting the applicability of the VT-2POD
method.

The VT-2POD isosceles triangle configuration may be inappropriate due to
differing centers of mass between the horizontal and vertical stimulus
configurations. A cross (\textbf{+}) configuration could ensure a consistent
center of mass, maintaining the validity of two-point spatial acuity
measurements as point separation approaches zero. However, the cross
configuration with four stimulus tips introduces even more challenges in
maintaining uniform contact pressure.

\subsubsection{Comparison with existing Experiments}
\label{sec5.1.2}
Our developed VT-2PD experiment has demonstrated potential validity for
generating information on vibrotactile perception in future studies. Existing
study designs typically allow for the investigation of one specific target skin
area (\cite{07_Johannesson.2017, 06_Hoffmann.2018, 08_Jouybari.2021}). Although we
have currently tested our experiment only on the forearm, our experimental
apparatus enables versatile application to different skin areas, thereby
facilitating an overall comparison of skin sensitivities. Test runs in other
skin areas have shown promising results and require formal validation in larger
studies.

The use of Bayesian adaptive parameter estimation (BAPE) in combination with
entropy minimization accelerates the procedure. The participant's response is
used to search for distances with the highest information content, saving
redundant queries and time. Consequently, more participants and skin areas can
be examined within a single test session.

\subsection{Discussion EPCE}
\label{sec5.2}

In our investigation of the Extended Perception Comparison Experiments (EPCE),
we found that the elevated standard deviation could be attributed to the
amalgamation of horizontal and vertical measurements into a single psychometric
function. In particular, the inclusion of vertical measurements appears to
significantly deteriorate the perception curve. This suggests that, especially
in the case of vibration, horizontal and vertical measurements should be
examined separately. The results hint at a strong anisotropy of perception,
which could be a crucial factor in future studies. While an S-shaped curve is
discernible, it exhibits a very flat progression.

\subsubsection{Comparison bidirectional VT-2PD and 2POD}
\label{sec5.2.1}

\begin{figure}[!t] 
  \centering
  \includegraphics[width=\figwidth]{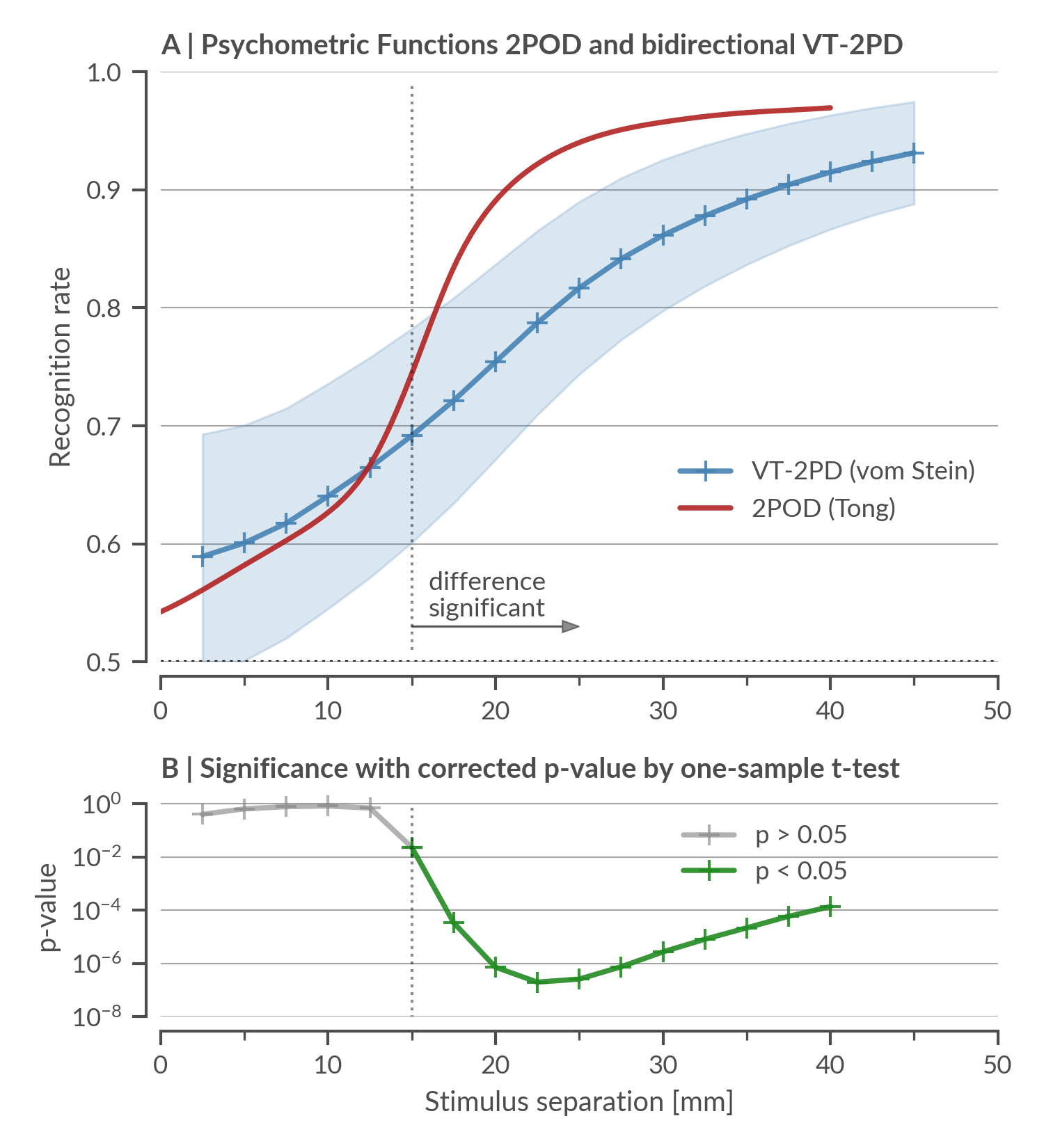}
  \caption{\textbf{EPCE}: Mean psychometric functions of bidirectional VT-2PD
  compared with 2POD by Tong et al. Recognition rate scale starts at $0.5$ for
  detailed view. Bonferroni corrected p-values of one-sample t-test are
  displayed on a logarithmic scale. Plus symbols (+) indicate the separations
  where measurements were taken.}
  \label{fig_comp_tong_vst}
\end{figure}

In order to compare our results with those of Tong et al.'s, we applied the
VT-2PD method bidirectionally, aligning our approach with theirs. The
psychometric functions of Tong et al. and ours (cf. Ch.~\ref{sec4.2}) are
superimposed in Figure~\ref{fig_comp_tong_vst}~\textbf{(A)}. To validate the
divergence between the curves, we conducted a one-sample t-test at each interval
step (+), with Bonferroni corrected p-values, plotted logarithmically in the
lower diagram \textbf{(B)}.

The curves exhibit pronounced differences from a distance of $15~mm$ onwards.
The p-value (log-scale) robustly confirms the significance of these differences.
The recognition rate in Tong et al.'s transition area ascends significantly
steeper, indicating a clear distinction in the perception of static and
vibrotactile stimuli. For the direction-independent secure recognition of
vibrotactile stimuli, a larger distance is necessitated compared to static
stimuli. This observation substantiates our hypothesis $H$.

  

\begin{figure}[!t] 
  \centering
  \includegraphics[width=\figwidth]{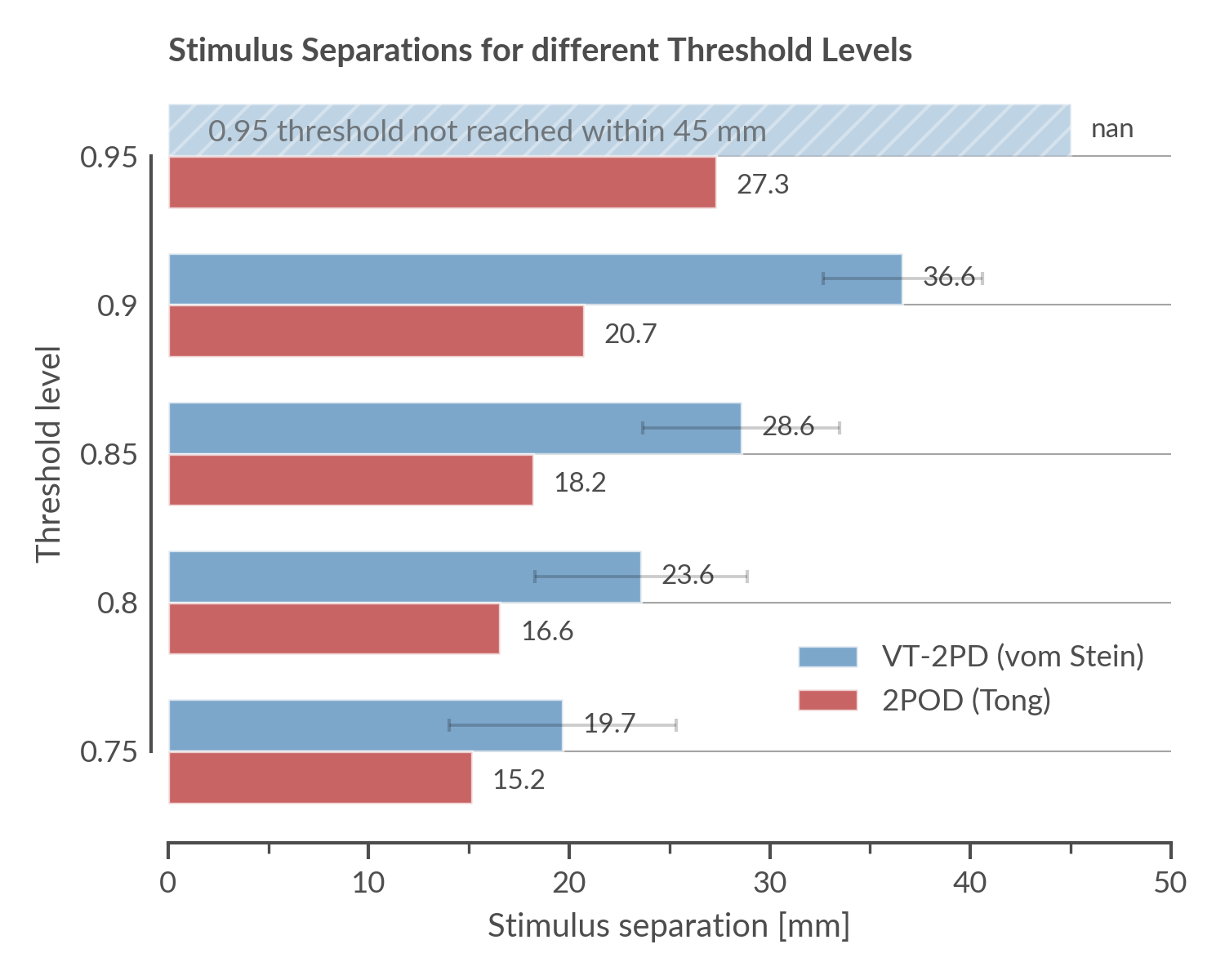}
  \caption{\textbf{EPCE}: Stimulus separation required to reach five different
  Threshold levels for vibrotactile stimuli (VT-2PD) and static stimuli (2POD,
  \cite{15_Tong.2013}). Error bars for VT-2PD: $\pm 1$ SE}
  \label{fig_thresholds_tong_vst}
\end{figure}

In Figure~\ref{fig_thresholds_tong_vst}, we compare the different stimulus
separations required for vibrotactile stimuli (VT-2PD) and static stimuli (2POD,
\cite{15_Tong.2013}) to reach five distinct threshold levels ($0.75$, $0.8$,
$0.85$, $0.9$, $0.95$) for both VT-2PD and 2POD. This comparison further
underscores the need for a significantly smaller stimulus separation in the 2POD
experiment to achieve the same threshold levels, ranging from $0.75$ to $0.9$,
while the $0.95$ level is not reached with VT-2PD within $45~mm$.

In aligning our study with that of Tong et al., we intentionally diverged in the
type of stimulus used, yet ensured a high degree of similarity in other aspects
of the methodology. Like Tong et al., we applied the stimuli bidirectionally,
thereby neglecting any anisotropy effects. Additionally, in our method, two
stimulus tips are always in contact with the skin, paralleling the conditions in
Tong et al.'s study. The use of the Bayesian Adaptive Parameter Estimation
(BAPE) algorithm in a manner akin to Tong et al.'s resulted in a comparable
question-answer paradigm. Moreover, our study was fully automated, which we
believe could further minimize any potential experimenter bias.

In terms of participant responses, we observed some notable differences between
our study and Tong et al.'s. These differences could provide valuable insights
into the perception of static and vibrotactile stimuli, and warrant further
investigation. For instance, Tong et al.'s study surpassed the $0.9$ recognition
rate threshold at $20.7~mm$, while ours did so at $36.6~mm$. Interestingly, the
$0.95$ threshold was not reached within our maximum distance of $45~mm$ with
vibrotactile stimuli, indicating potential limitations in vibrotactile
perception. This could be due to the propagation of vibrations in the skin
leading to reduced spatial resolution, as suggested by \cite{Shah.2019}.

This observation raises questions about the transferability of spatial acuity
values between static and vibrotactile stimuli, suggesting that vibrotactile
perception requires a completely new investigation. Whether a reliable $0.95$
level can be achieved with vibrotactile stimuli remains an open question.

Finally, our findings have important implications for the existing body of
knowledge. By directly comparing our results with those of Tong et al.'s, we are able
to provide new insights into the perception of vibrotactile stimuli. This not
only confirms our hypothesis but also contributes to a deeper understanding of
this complex phenomenon. The potential correlation between vibrotactile and
static perception, possibly represented by a transformation function, can only
be confirmed in further studies.

\subsection{Limitations}
\label{sec5.3}
While our study has made significant strides in the field of tactile perception
research, it is not without its limitations, which we acknowledge as
opportunities for future research and refinement of our methodology.

Firstly, to achieve the smallest possible stimulus separation and the highest
possible application accuracy, the stimulus application via the stimulus tips is
very small-scale, almost point-like. In tactile displays, larger application
areas have mostly been used so far (\cite{03_Cholewiak.2003, 07_Johannesson.2017,
02_Cholewiak.2004}). This might have an increased influence on spatial acuity
and requires further investigation.

Secondly, the automation of the apparatus eliminates human bias in the stimulus
application during the experiment. However, inadequate alignment of the
apparatus can render a run unusable. The alignment on the skin area before the
start of the experiment must be carried out with great care and monitored during
the experiment. Test runs exhibiting indications of bias should be rigorously
evaluated and, if necessary, excluded from the analysis (cf. Ch. \ref{sec3.3.2}).

Thirdly, although our apparatus offers high flexibility in its application, it
has limitations in examining certain skin areas. Uneven areas and regions with
variable skin thickness (subcutaneous fat) present challenges in ensuring
uniform pressure application across all stimulus separations, potentially
introducing bias. Areas with limited space, such as the hands and fingers, are
also less amenable to examination. For these smaller regions, however, a more
compact version of the experimental apparatus can be utilized.

In conclusion, while our study has provided valuable insights into the
perception of vibrotactile stimuli, these limitations highlight the need for
further research to refine our methodologies and expand the scope of our
investigations.

\subsection{Future Work}
\label{sec5.4}
Following the development of the experimental apparatus in this study, a
large-scale study with a minimum of 30 participants will be conducted. The
processes for the VT-2PD trial will be optimized based on the insights gained
from the preliminary study, and potential sources of error will be eliminated.\\
For the implementation of this study, we propose two hypotheses:

\vspace{1em}
\noindent\textbf{$H_1$}: \textit{Sensitivity to vibrotactile stimuli varies
significantly across different skin areas of the body.}

\vspace{1em}
\noindent\textbf{$H_2$}: \textit{Perception of vibrotactile stimuli on the skin
is anisotropic - exhibiting significant differences in various orientations at
the same skin area.}
\vspace{1em}

We aim to establish direct comparability of vibrotactile sensitivity across
different skin areas by investigating $H_1$. To this end, we will conduct our
study on the forearm, lower back, abdomen, and thigh areas, and evaluate the
results using statistical analyses.

We plan to investigate $H_2$ by conducting the experiment in both horizontal
and vertical directions for each skin area, thereby establishing separate
psychometric functions, this time.

In addition to the large-scale study and the exploration of the hypotheses
mentioned, the apparatus developed in this research presents potential for
further adaptations and applications. One such possibility, without committing
to a definite plan, is the adaptation of the apparatus for electrostimulation
studies. The design of our apparatus allows for such modifications with minimal
effort, opening up new avenues for research in tactile perception. However, the
feasibility and implications of this adaptation would need to be thoroughly
evaluated in the context of future work.

The primary objective of the subsequent investigation is to contribute to the
design principles of tactile displays. This includes focusing on the
dimensioning of tactile interfaces, determining the appropriate spacing of
actuators with respect to the skin area, and considering the spacing of
actuators in relation to the directional vector on the skin
(horizontal/vertical).

\section{Conclusion}
\label{sec6}
This study has made substantial advancements in the field of tactile perception
research by 1) introducing a verified novel and versatile approach for measuring
the spatial acuity of vibrotactile stimuli and 2) enabling a direct comparison
between vibrotactile and static stimuli on the forearm. Our findings have
confirmed the hypothesis that the spatial acuity of human skin for vibrotactile
stimuli is significantly different from that for static stimuli. This discovery
has profound implications for the design and application of tactile interfaces,
particularly in the context of assistive technologies for visually impaired
individuals.

In part 1) of our research, we developed and evaluated two distinct methods
for conducting vibrotactile experiments: the vibrotactile
Two-Point-Discrimination (VT-2PD) and the vibrotactile
Two-Point-Orientation-Discrimination (VT-2POD). The VT-2PD emerged as a robust
and reliable method for conducting initial experiments, demonstrating its
effectiveness and potential for future studies. This method allows for the
investigation of anisotropy effects in future studies and can be applied to
uneven skin areas, offering a broad spectrum of investigation possibilities.
Furthermore, the use of the Bayesian Adaptive Parameter Estimation (BAPE)
algorithm in conjunction with the VT-2PD method allows for significant results
with fewer trials by actively searching for threshold values rather than
relying on fixed interval steps. In contrast, the VT-2POD did not yield
satisfactory results and will therefore be discontinued. 

In part 2) of our research, we have successfully conducted a comprehensive study
with 23 participants, enabling us to directly compare our findings with those of
Tong et al. This confirmed our hypothesis of significant differences in stimulus
separation from a distance of $15~mm$ upwards. Our results indicate that
vibrotactile perception exhibits lower acuity at higher separations, potentially
due to the propagation of vibrations through the skin, thereby penetrating
additional receptive fields. This observation underscores the fact that
guidelines for the perception of static stimuli (2PD, 2POD) cannot be directly
applied to vibrotactile perception.

Our findings provide a compelling argument for the need to consider the unique
characteristics of vibrotactile stimuli when designing and interpreting tactile
perception experiments. The potential influence of vibration propagation on
spatial acuity is a complex issue that warrants further investigation.

Looking forward, we plan to conduct a large-scale study with a minimum of 30
participants to further investigate the variability of vibrotactile sensitivity
across different skin areas and the potential anisotropy of vibrotactile
perception. The insights gained from this future work will be invaluable in
informing the design principles of tactile displays, including the dimensioning
of tactile interfaces, the appropriate spacing of actuators with respect to the
skin area, and the orientation of actuators on the skin.

In conclusion, our research has provided a new perspective on the perception of
vibrotactile stimuli, offering valuable insights that will guide future research
and the development of tactile interfaces. We look forward to the continued
exploration of this fascinating aspect of human sensory perception.

\section*{Declaration of Competing Interest}
The authors declare that they have no known competing financial interests or
personal relationships that could have appeared to influence the work reported
in this paper.


\section*{Acknowledgments}
The authors express gratitude to our university for funding this research, the
voluntary test subjects for their participation in the pilot study, and members
of the science community across multiple disciplines for their valuable
feedback. We appreciate the support and contributions from all parties involved.

\section*{Author Contributions}
\textit{vom Stein}: Conceived and led the research topic, developed the concept,
constructed and built the test apparatus, programmed the experimental setup,
conducted the central part of the research and played a leading role in the
overall investigation, and wrote the majority of the manuscript along with
creating the figures.

\textit{Hoppe}: Conducted research, primarily contributed to the writing of the
mathematical section of the paper, consulted on technical difficulties with the
apparatus, collaborated on initial experiments, and participated in discussions
for the development of the study design.

\textit{Sommer}: Conducted research, contributed to the writing of the introduction
section of the manuscript, collaborated on initial experiments, and participated
in discussions for the development of the study design.

\textit{Wolf}: Provided funding through the university, offered guidance and
consultation, and reviewed the manuscript.

\bibliographystyle{unsrtnat}
\bibliography{2023_vomStein}  






\end{document}